# Molecular design and control of fullerene-based bi-thermoelectric materials

Laura Rincón-García[1,2], Ali K. Ismael[3,4], Charalambos Evangeli[1], Iain Grace[3], Gabino Rubio-Bollinger[1,5], Kyriakos Porfyrakis[6], Nicolás Agraït[1,2,5,*] and Colin J. Lambert[3,*]

[1]*Departamento de Física de la Materia Condensada and Condensed Matter Physics Center (IFIMAC), Universidad Autónoma de Madrid, E-28049 Madrid, Spain*

[2]*Instituto Madrileño de Estudios Avanzados en Nanociencia IMDEA-Nanociencia, E-28049 Madrid, Spain*

[3]*Department of Physics, Lancaster University, Lancaster, United Kingdom*

[4]*Department of Physics, College of Education for Pure Science, Tikrit University, Tikrit, Iraq*

[5]*Instituto Universitario de Ciencia de Materiales "Nicolás Cabrera", Universidad Autónoma de Madrid, E-28049 Madrid, Spain*

[6]*Department of Materials, University of Oxford, Oxford OX1 3PH, United Kingdom*

*Corresponding authors: nicolas.agrait@uam.es, c.lambert@lancaster.ac.uk

**Molecular junctions are a versatile test bed for investigating thermoelectricity on the nanoscale[1-10] and contribute to the design of new cost-effective environmentally-friendly organic thermoelectric materials[11]. It has been suggested that transport resonances associated with the discrete molecular levels would play a key role in the thermoelectric performance[12,13], but no direct experimental evidence has been reported. Here we study single-molecule junctions of the endohedral fullerene $Sc_3N@C_{80}$ connected to gold electrodes using a scanning tunnelling microscope (STM). We find that the magnitude and sign of the thermopower depend strongly on the orientation of the molecule and on applied pressure. Our theoretical calculations show that the $Sc_3N$ inside the fullerene cage creates a sharp resonance near the Fermi level, whose energetic location and hence the thermopower can be tuned by applying pressure. These results reveal that $Sc_3N@C_{80}$ is a bi-thermoelectric material, exhibiting both positive and negative thermopower, and provide an unambiguous demonstration of the importance of transport resonances in molecular junctions.**

**Keywords**: thermopower, molecular electronics, scanning tunnelling microscopy

The design of new thermoelectric materials for converting waste heat directly into electricity is a global challenge[14]. A successful strategy for improving thermoelectric properties of inorganic semiconductors is nanostructuring, which leads to quantum confinement of electrons and suppression of parasitic phonons[15]. However, the cost and toxicity of currently-available inorganic thermoelectric devices are problematic and make organic thermoelectrics particularly attractive for sustainable waste energy conversion[11]. Recently, the ability to measure thermoelectric properties of single molecules has begun to deliver the fundamental physical and chemical knowledge, which will be needed to design new materials and build devices suitable for applications. It has been demonstrated both experimentally and theoretically that, at a molecular scale, thermopower $S$ can be controlled by varying the chemical composition[1], varying the position of intra-molecular energy levels relative to the work function of metallic electrodes[2,3], systematically increasing the single-molecule lengths and varying the binding groups within a family of molecules[4-8], by tuning the interaction between two neighbouring molecules[9], and by controlling the transport properties with an electrostatic gate[10] or electrochemically[16]. These single-molecule experiments yielded room-temperature values of $S$ ranging in magnitude from ca. 1 to 50 µV/K.

A key factor in determining the thermoelectric performance of a molecular junction, according to theoretical calculations, is the presence of transport resonances close to the Fermi level. The effect of transmission resonances was explored theoretically by Finch et al.[12]. In this work the resonances were introduced by quantum interference effects due to the addition of side groups to the molecular backbone. The resulting Fano resonance near the Fermi level leads to a large enhancement in the thermopower and thermoelectric efficiency. Enhanced thermoelectric properties have also been predicted in the vicinity of interference-induced transmission nodes[13]. However, this approach remains experimentally unexplored.

Our recent experiments and theory of $C_{60}$-based thermoelectricity[9] and other studies[2,3,10] showed that $C_{60}$ is a robust thermoelectric material, with a consistently-negative thermopower, which originates from the presence of a broadened LUMO level near the Fermi energy $E_F$ whose monotonic shape makes it difficult to vary the sign of $S$. Therefore we reasoned that if a more narrow transport resonance could be created within the tail of the LUMO, this could lead to increased tunability. To create such a feature, we selected the endohedral fullerene $Sc_3N@C_{80}$, which is known to have a relatively small energy gap[17] and to be particularly stable at room temperature and even at elevated temperatures[18]. As shown below, the presence of the $Sc_3N$ inside the fullerene cage not only creates the desired resonance, but also, depending on the orientation of the molecule, allows tuning of the position of the resonance and hence the sign of the thermopower by mechanically compressing the junction. This experimental tuning of the sign of $S$ by mechanical gating is unprecedented and is the first example of a bi-thermoelectric material, which can exhibit thermopower of either sign, without doping and without a change of chemical composition. Mechanical modulation of the thermopower of molecular junctions has been theoretically predicted[19], but as we will show the origin of this effect is different.

In this Letter, we use a modified scanning tunnelling microscope (STM)[9] to study single-molecule junctions of an endohedral fullerene $Sc_3N@C_{80}$ between gold electrodes. $Sc_3N@C_{80}$ consists of a fullerene cage ($C_{80}$-Ih) encapsulating three scandium atoms joined to one nitrogen, as shown schematically in Fig. 1a. Supplementary Fig. 1 and Methods contain more information about molecule purification and deposition. All STM experiments were performed at room temperature and in ambient conditions. STM imaging after deposition shows isolated molecules as well as aggregates at step edges or in small islands (Fig. 1b-d). Tunnelling spectroscopy on different isolated molecules reveals an asymmetric current-voltage characteristic (rectifying behaviour). We found that some molecules have higher conductance for positive voltage (black curve in Fig. 1e-f), similar to what is observed in $C_{60}$, while others have higher conductance for negative voltage (pink curve in Fig. 1e-f). This variation in the electronic behaviour of particular molecules of $Sc_3N@C_{80}$ has been reported in UHV experiments[18] and has been attributed to the orientation of the molecule on the surface.

Next we measured simultaneously the conductance $G$ and thermopower $S$ of isolated $Sc_3N@C_{80}$ molecules (usually sitting on a step edge) as described in ref. 9 (Fig. 2c) (see also Methods and Supplementary Fig. 2 & 3). Fig. 2a-b show examples of $G$ (in blue) and $S$ (in green) curves measured on two different junctions while the tip approaches to touch the molecule. We found that the conductance of $Sc_3N@C_{80}$ behaves similarly to the case of $C_{60}$ junctions with a jump-to-contact signalling the first-contact as the tip atoms touch the molecule. Typical values for the first-contact conductance are smaller by a factor of three when compared to $C_{60}$ (as shown in Fig. 2d). The thermopower at first-contact of $Sc_3N@C_{80}$ molecules, in contrast to $C_{60}$, can be positive, negative, or close to zero depending on the selected molecule (Fig. 2a-b), resulting in a broad histogram centred around zero, as shown in Fig. 2e.

Interestingly, we find a correlation with the tunnelling spectroscopy results: molecules for which the conductance is higher at positive voltage show negative thermopower at first-contact, whereas molecules for which the conductance is higher at negative voltage show positive thermopower. This is consistent with the fact that the thermoelectric properties of a molecular junction depend on the magnitude and derivative of the transmission at the Fermi level of the electrodes,[20]

$$S = -\frac{\pi^2 k_B^2 T}{3e} \frac{d \ln \mathcal{T}(E)}{dE}\bigg|_{E=E_F} \quad (1)$$

where $\mathcal{T}(E)$ is the junction transmission, which is dependent on electron energy $E$, $k_B$ is the Boltzmann constant, $T$ is the average temperature of the junction, and $e$ is the charge of the electron.

We now investigate the variation of thermopower as the tip advances after the first contact, compressing the $Sc_3N@C_{80}$ molecule. We positioned the tip on a selected isolated molecule and performed small amplitude (< 0.5 nm) compression (approach-retraction) cycles, always maintaining contact with the molecule (see Supplementary Fig. 4). In these cycles a variable pressure is exerted on the molecule by the STM tip. In Fig. 3a-b, we present the simultaneous variations of $G$ and $S$ measured during three cycles for three different molecules. The periodic nature of these curves indicates that, in response to the pressure, the junction, i.e., the molecule and the gold electrodes, deforms elastically. Larger amplitude (> 0.5 nm) cycles destroy this periodicity indicating the onset of plastic deformation (atomic rearrangements) in the gold electrodes[21] (see Supplementary Fig. 5 for more details). Taking into account previous results for gold contacts[22] and the fact that fullerene molecules are much stiffer than gold[23], we can safely assume that most of the elastic deformation corresponds to the electrodes and that the maximum pressure at the junction during our measurements is about 4 GPa[21].

The traces shown in Fig. 3a-c correspond to three molecules with different behaviours: the red traces correspond to a molecule which showed large positive thermopower at first contact (molecule 1); the blue traces, to a molecule with small positive thermopower (molecule 2); and the green traces, to a molecule with almost zero thermopower (molecule 3). We observe that for all molecules both the conductance and thermopower vary monotonically with pressure:

the conductance increases and the thermopower decreases, becoming more negative, as the tip presses the molecule. This behaviour of the conductance is to be expected, since pressing will result in an increased coupling and consequently in a larger conductance. However, the behaviour of the thermopower is most unusual: very large variations are observed and even a change in sign for molecule 2. This extreme sensitivity of thermopower of molecular junctions to pressure had never been reported and has a marked effect on the power factor, $GS^2$, as shown in Fig. 3c. For molecule 1, the power factor decreases with compression, while for molecule 2, it increases reaching values of around 5 fW/K$^2$. In contrast, for molecule 3, $GS^2$ remains small during the whole cycle.

To elucidate the origin of the bi-thermoelectric effect of the endohedral fullerene junctions we use density functional theory (DFT) to simulate the contact and pressing of the Sc$_3$N@C$_{80}$ molecule. Using a combination of the quantum transport code Gollum[24] and the DFT code SIESTA[25], we calculate both the conductance and thermoelectric properties of the molecule contacted between gold electrodes (a detailed description can be found in the Supporting Information). Three major inputs enter into the simulations. The first is the position of the molecule with respect to the electrodes $z$. We explore the effect of pressure on the transport properties by varying $z$ around the equilibrium distance, which is found to be approximately 2.3 Å (see Supplementary Fig. 6). The orientation of the whole molecule is kept fixed while $z$ is varied. A second input to the simulations is the orientation of the molecule with respect to the gold surface. The electronic structure of the isolated molecule shows a LUMO resonance located primarily on the Sc$_3$N molecule (see Supplementary Fig. 7) and therefore transport properties are expected to depend on the orientation of the Sc$_3$N, locked in position within the fullerene cage (see Supplementary Fig. 8), with respect to the gold surface. We define $\theta = 0º$ to be the orientation when the plane of the Sc$_3$N molecule is normal to the gold surface, such that at $\theta = 90º$ the Sc$_3$N is parallel to the surface (see Supplementary Fig. 6). To explore a range of orientations, we rotate through 180º, at intervals of 3º, and at each angle compute the zero bias transmission coefficient $\mathcal{T}(E)$ (Supplementary Fig. 9). A third important input to the simulations is the energetic location of the molecular energy levels relative to the Fermi energy of the electrodes, since the conductance and thermopower are related to the value of the transmission and its derivative at the Fermi energy. Since the DFT-predicted value $E_F^0$ is not reliable, to find the *true* Fermi level, we compute the transmission as a function of the energy for different orientations and pressures (see Fig. 3g-i and Supplementary Figs. 9 and 10) and identify a single Fermi energy that reproduces the experimentally observed behaviour, i.e., the thermopower is either positive or negative and it always decreases shifting to more negative values as the molecule is pressed, in some cases passing through zero. We find that the *true* Fermi level is located between the LUMO resonance and the LUMO and takes a value $E_F = E_F^0 + 0.23\ eV$.

With this choice of $E_F$, Fig. 3d-f show $G$, $S$ and $GS^2$ as a function of $z$, for three different orientations of the endohedral molecule ($\theta = 150º$ in red, $\theta = 57º$ in blue, and $\theta = 63º$ in green), which match the experimental behaviour of molecules 1, 2 and 3 in Fig. 3a-c. These results illustrate that the diverse experimental behaviours can be attributed to different orientations of the endohedral fullerene and reflect the shift towards negative values of the thermopower with pressure. The origin of this effect lies in the extreme sensitivity to pressure and orientation of the transmission function $\mathcal{T}(E)$, due to the presence of the resonance close to the Femi level (Fig. 3g-i). For all the orientations (Supplementary Fig. 11), as the tip advances, the resonance becomes broadened and shifts to lower energies, as a consequence of changes in the imaginary and real parts, respectively, of the self-energy[26], that is, in the coupling of the molecule to the electrodes. This results in an increase of the value of the conductance while the value of the thermopower becomes more negative as the molecule is pressed. For certain orientations, the junction shows a small positive thermopower and pressing the molecule produces an $S$ that varies from positive to negative values (Fig. 3h), in good agreement with experimental results.

To further illustrate formation of the LUMO resonance due to the $Sc_3N$, we have performed the same experiments on $C_{60}$ junctions (Fig. 4a-c). We find that the thermopower is negative during the whole cycle and the power factor increases with compression to values ~ 10 fW/K$^2$. Theoretical calculations show that for C$_{60}$ junctions the Fermi level is always located in the smooth, increasing slope of the LUMO peak (see Fig. 4d-g). Figures 4i and 4h show the transmission functions of Sc$_3$N@C$_{80}$ and C$_{60}$ junctions, respectively, clearly showing that the main difference between these two systems is the LUMO resonance in the case of the endohedral fullerene.

Our calculations reveal the essential role played by the coupling of the molecule to the electrodes in the observed changes in the thermopower, while the deformation of the molecule plays only a minor role, in contrast with the mechanism proposed in ref. 19, which relied in intramolecular deformation.

In conclusion, we have demonstrated a new concept of bi-thermoelectricity, in which the sign and magnitude of the thermopower of a given material can be tuned. This effect was realised by identifying a molecule with a transmission resonance close to the Fermi energy, whose energetic location is sensitive to orientation and pressure. In this paper, we demonstrated bi-thermoelectricity in Sc$_3$N@C$_{80}$, but more generally the effect should be present in any material with orientation-dependent and pressure-dependent transmission resonances, which can be caused to pass through the Fermi energy. For the future, if appropriate templating strategies could be implemented, which select appropriate orientations of such molecules, then single-material, nanoscale tandem devices with alternating-sign thermopowers could be realised.

## Methods

**STM measurements.** The gold surfaces were flamed annealed before endohedral fullerenes were deposited, using the drop casting, technique from solution in 1,2,4-trichlorobenzene (TCB) at very low concentration (10$^{-7}$-10$^{-8}$ M). We used mechanically cut gold tips as STM probes. In order to measure the thermopower of molecular junctions, we modified our homebuilt STM setup by adding a surface mount 1 kΩ resistor that acts as a heater for the tip holder[9], while the substrate was kept at room temperature. Two thermocouples connected to the tip holder and to the substrate surface were used to monitor the resulting temperature difference. Measurements were performed at a temperature difference of 40 K and at ambient conditions. We found that the temperature stabilizes in about 15-30 minutes. The error in the determination of the thermopower is of about 0.6 µV/K. Further details are provided in the Supplementary Information.

**Computational details.** Electronic structure calculations were performed using the DFT code SIESTA[25]. The optimum geometry of the $C_{80}$ cage and encapsulated $Sc_3N$ molecule was obtained by searching through atomic configurations until the lowest energy was found. In each case the molecule was relaxed until all the forces on the atoms were less than 0.05 V/Ang. SIESTA employs pseudo-atomic orbitals and the relaxation was carried out using a double-zeta plus polarization orbital basis set. Norm-conserving pseudopotentials were used and an energy cutoff of 200 Rydergs defined the real space grid. The exchange correlation functional was LDA. To calculate electron transport, the molecule was attached to gold leads, and due to the large contact area of the $C_{80}$ cage, a 5 by 6 atom layer of (111) gold was taken to be the surface of the lead. The optimum binding location was found by calculating the binding energy as a function of the separation distance $z$ taking into account basis set error corrections. Supplementary Fig. 6 shows that this distance is approximately 2.3 Angstroms. The extended molecule included 6 layers of (111) gold and the Hamiltonian describing this structure was produced using SIESTA. The transmission coefficient $\mathcal{T}(E)$ and thermopower $S$ were calculated using the Gollum code[24]. Further details are provided in the Supplementary Information.

## Acknowledgements

This work was supported by the Swiss National Science Foundation (No. 200021-147143) as well as by the European Commission (EC) FP7 ITN "MOLESCO" project no. 606728, UK EPSRC, (grant nos. EP/K001507/1, EP/J014753/1, EP/H035818/1), Spanish MINECO (grant nos. MAT2011-25046 and MAT2014-57915-R) and Comunidad de Madrid NANOFRONTMAG-CM (S2013/MIT-2850) and MAD2D-CM (S2013/MIT-3007), and the Iraqi Ministry of Higher Education, Tikrit University (SL-20). L.R.-G. acknowledges financial support from UAM and IMDEA-Nanoscience. A. K. I. acknowledges financial support from Tikrit University.

## Author contributions

L.R.-G. performed the experiments and analysed the experimental data. A.K.I. and I.G. carried out the theoretical calculations. K.P. purified and provided the endohedral molecules. C.E. contributed to the experiments and to the experimental setup and G.R.-B. contributed to the experimental setup. N.A. and C.J.L. conceived and supervised the experiment and wrote the manuscript with contributions from all the authors.

## Additional information

Supplementary information containing details on experimental and theoretical procedures is available in the online version of the paper. Reprints and permission information is available online at http://www.nature.com/reprints. Correspondence and requests for materials should be addressed to N.A. or C.J.L.

## Competing financial interests

The authors declare no competing financial interests.

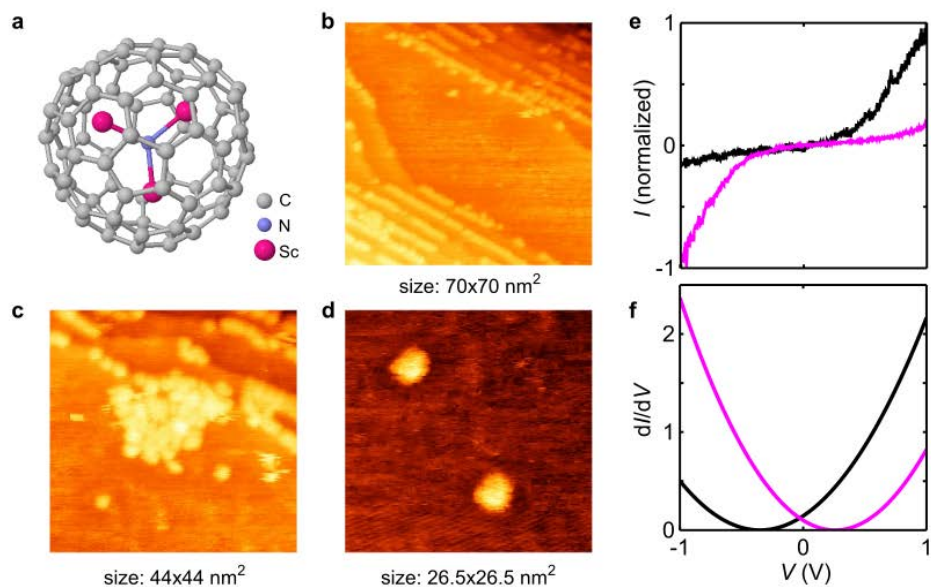

**Figure 1 | Scanning tunnelling microscope images and tunnelling spectroscopy. a**, Schematic of the endohedral fullerene $Sc_3N@C_{80}$ used in this work; notice the $Sc_3N$ in the centre of the fullerene cage. **b-d** STM images of the molecules on atomically flat (111) Au surfaces showing preferential adsorption at step edges (**b**), islands (**c**) and isolated molecules (**d**). **e,f** IV characteristics (**e**) and differential conductance (**f**) in tunnelling regime on two different molecules, presenting opposite rectifying behaviour.

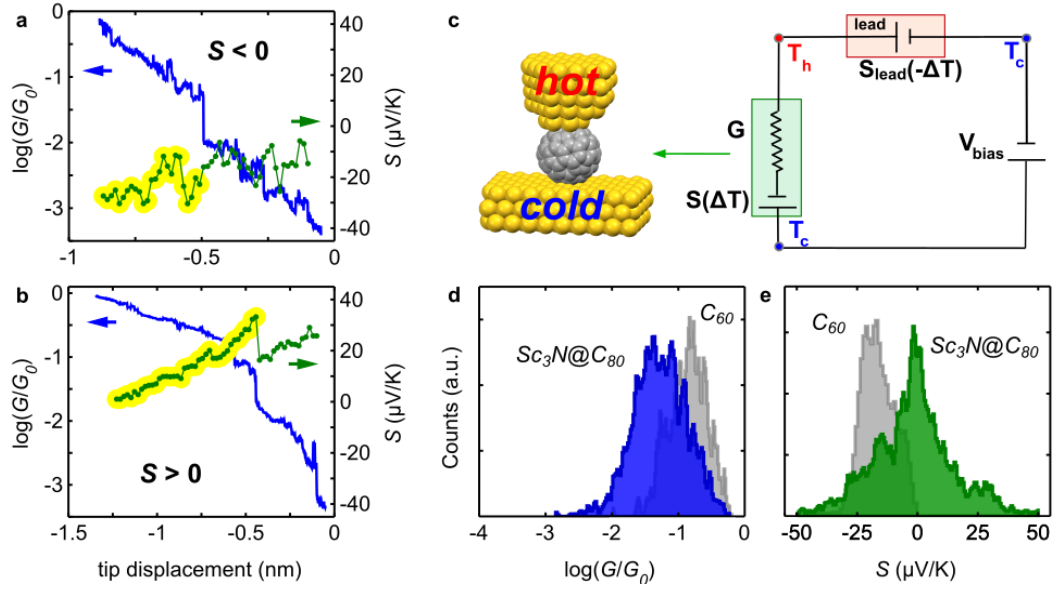

**Figure 2 | Thermopower and conductance simultaneous measurements. a,b**, Conductance, $G$ (blue), and thermopower, $S$ (green), simultaneously acquired while approaching individual Sc$_3$N@C$_{80}$ molecules. For the molecule in **a**, the thermopower is always negative, while for that one in **b**, it is always positive. In these measurements the temperature difference was $\Delta T \simeq 40$ K. $G_0 = 2e^2/h$, where e is the electron charge, and $h$ is Panck's constant, is the conductance quantum. The portion of the thermopower trace highlighted in yellow corresponds to molecular contact with the tip. **c**, Schematic representation of the experimental setup. The tip is heated to a temperature $T_h$ above ambient temperature $T_c$, while the substrate is maintained at $T_c$ (see Supplementary Fig. 3 for more details of the thermal circuit). **d,e**, Conductance $G$ and thermopower $S$ at first-contact histograms of Sc$_3$N@C$_{80}$ (in blue and green, respectively) compared to C$_{60}$ (in grey). For the Sc$_3$N@C$_{80}$ histograms, the mean conductance value is $\bar{G} = 0.05\ G_0$ and the mean thermopower is $\bar{S} = -2$ μV/K.

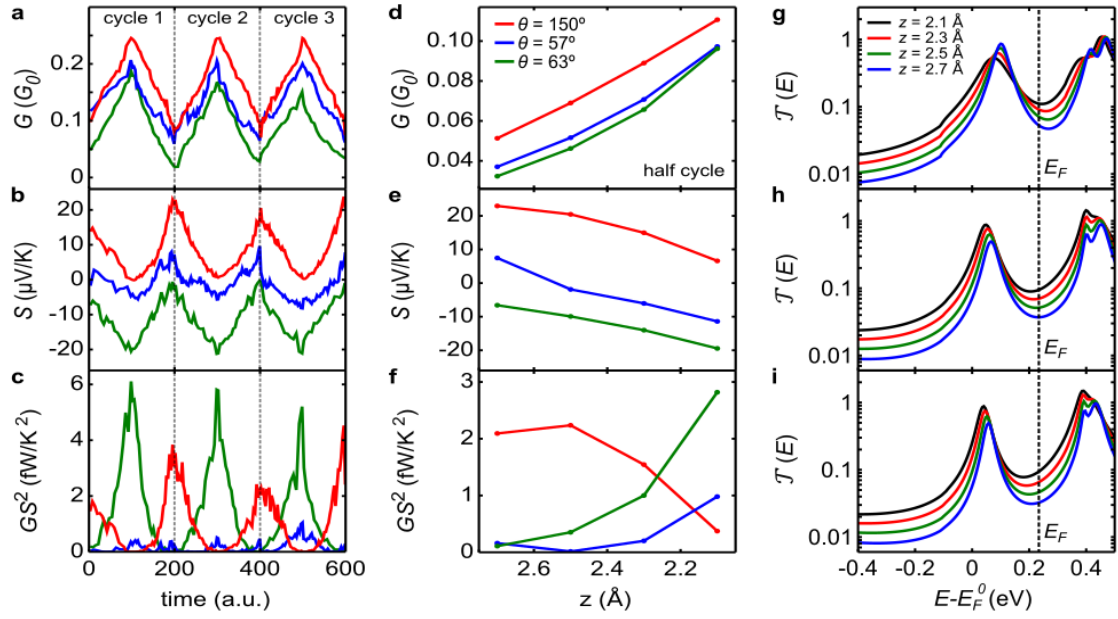

**Figure 3 | Effect of pressure on Sc$_3$N@C$_{80}$ molecular junctions. a-c,** Periodical variations of the conductance $G$, thermopower $S$ and power factor $GS^2$, respectively, as the STM tip advances and retracts during three cycles. Each half cycle corresponds to less than 0.5 nm. Each colour corresponds to a different molecule. **d-f,** Calculated $G$, $S$ and $GS^2$, respectively, for three different orientations $\theta$ as the tip-molecule separation $z$ decreases from 2.7 Å to 2.1 Å, which corresponds to increasing pressure in the first half of experimental cycles. The orientations have been chosen such as to present similar amplitude variations as the experimental curves. **g-i,** Transmission curves, $\mathcal{T}(E)$, for the same three different orientations ($\theta = 150º$, $\theta = 57º$, and $\theta = 63º$, respectively) and for different $z$. The Fermi level is shifted from the position given by DFT and the black dotted line indicates the *true* Fermi level as explained in the text (see also Supplementary Fig. 9 and 10).

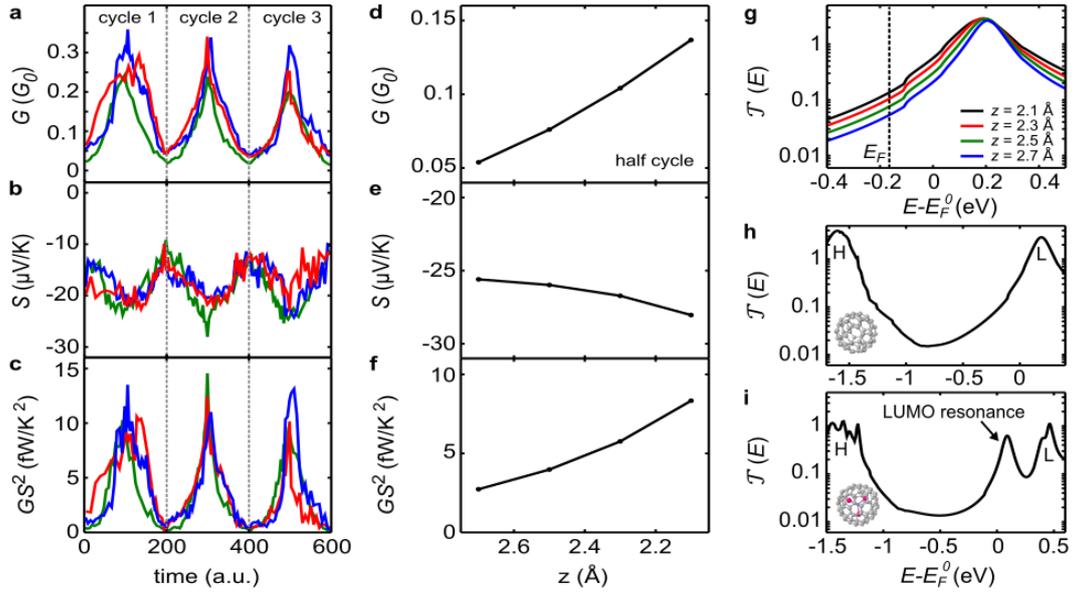

**Figure 4 | Effect of pressure on $C_{60}$ molecular junctions and comparison with $Sc_3N@C_{80}$ junctions. a-c,** Periodical variations of the conductance $G$, thermopower $S$ and power factor $GS^2$ of $C_{60}$ respectively, as the STM tip advances and retracts during three cycles. Each colour corresponds to a different molecule. Each half cycle corresponds to less than 0.5 nm. **d-f,** Calculated $G$, $S$ and $GS^2$, respectively, as the tip-molecule separation $z$ decreases from 2.7 Å to 2.1 Å, which corresponds to increasing pressure in the first half of experimental cycles. **g,** Transmission curves, $\mathcal{T}(E)$, for different $z$. The Fermi level is shifted from the position given by DFT and the black dotted line indicates the *true* Fermi level as explained in the text. In this case, $E_F = E_F^0 - 0.165\ eV$, chosen such as to present similar amplitude variations as the experimental curves. **h,i,** Transmission curves, $\mathcal{T}(E)$, for $C_{60}$ and $Sc_3N@C_{80}$ junctions, respectively. Letters H and L indicate the HOMO and LUMO peaks of the fullerene cages. The main difference between both systems is the resonance present in the case of the endohedral fullerene.

# Molecular design and control of fullerene-based bi-thermoelectric materials

# Supplementary Information

**Laura Rincón-García[1,2], Ali K. Ismael[3,4], Charalambos Evangeli[1], Iain Grace[3], Gabino Rubio-Bollinger[1,5], Kyriakos Porfyrakis[6], Nicolás Agraït[1,2,5,*] and Colin J. Lambert[3,*]**

[1]*Departamento de Física de la Materia Condensada and Condensed Matter Physics Center (IFIMAC), Universidad Autónoma de Madrid, E-28049 Madrid, Spain*

[2]*Instituto Madrileño de Estudios Avanzados en Nanociencia IMDEA-Nanociencia, E-28049 Madrid, Spain*

[3]*Department of Physics, Lancaster University, Lancaster, United Kingdom*

[4]*Department of Physics, College of Education for Pure Science, Tikrit University, Tikrit, Iraq*

[5]*Instituto Universitario de Ciencia de Materiales "Nicolás Cabrera", Universidad Autónoma de Madrid, E-28049 Madrid, Spain*

[6]*Department of Materials, University of Oxford, Oxford OX1 3PH, United Kingdom*

*Corresponding authors: nicolas.agrait@uam.es, c.lambert@lancaster.ac.uk

**Table of Contents**

# 1. Molecular purification details

Raw material was purchased from SES Research and by high-performance liquid chromatography (HPLC) was further purified to more than 99%, as shown by the HPLC trace in Fig. S1.

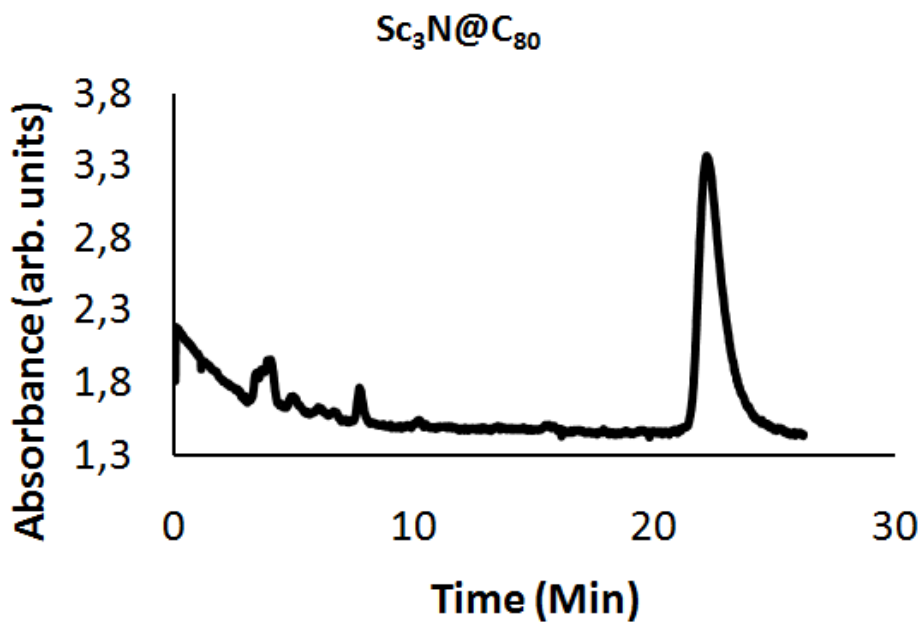

**Figure S1 | HPLC trace for the purification of $Sc_3N@C_{80}$.** HPLC equipment supplied by JAI. HPLC column supplied by Nacalai Tesque. Eluent: pure toluene, flowrate: 16 ml/min, column: Buckyprep-M (20 x 250 mm). $Sc_3N@C_{80}$ elutes between 20 and 25 mins.

# 2. Substrate preparation, experimental setup and technique

Endohedral fullerenes are deposited using the drop casting technique from a very dilute ($10^{-7}$ -$10^{-8}$ M) 1,2,4-trichlorobenzene solution. Specifically, a drop of the solution is left on an annealed gold surface for about 3 minutes, and then is blown off with dry nitrogen and allowed to dry overnight. Once the sample is dry, we mount it on a homebuilt STM and let it stabilize for about one hour in order to minimize the thermal drift. Using this procedure we are able to deposit isolated molecules both on terraces and step edges and forming small islands, as shown in Fig. 1b-d in the main text. The molecules at the steps are generally more stable under scanning.

In order to measure the thermopower of the molecular junctions, we have modified our STM setup by adding a surface mount 1 kΩ resistor which acts as a heater to the tip holder while the substrate was maintained at room temperature. Two thermocouples connected to the tip and sample holders were used to monitor the resulting temperature difference, which was set to approximately 40 K in the herein reported experiments; the sample is maintained approximately at 298 K (room temperature). We found that the temperature stabilized in about 15-30 minutes and the thermal drift increased, making necessary to use fast imaging to locate the isolated molecules.

The thermopower of molecular junctions is measured during the approach and the retraction of the STM tip to the molecule as described in Fig. S2.

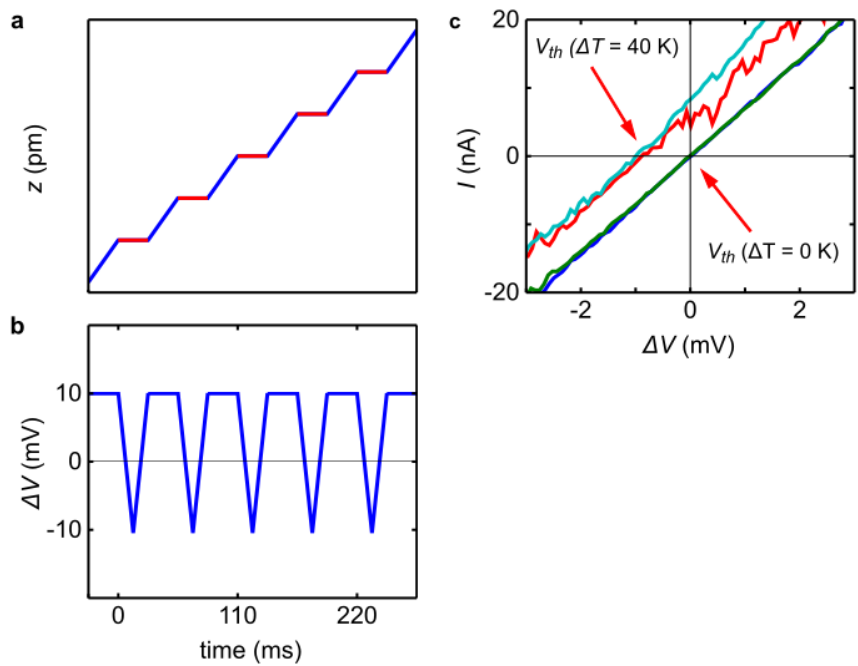

**Figure S2 | Technique for the simultaneous measurement of conductance and thermopower**. **a,b,** Tip displacement $z$ and applied bias voltage $\Delta V$ at the molecular junction, respectively, as a function of time. The bias voltage is maintained at a fixed value $\Delta V_0$ during the tip motion (in blue in **a**) and every few picometers it is swept between $\pm \Delta V_0$ while the tip is stationary (in red in **a**). In the experiments, the bias voltage $\Delta V_0$ was set to 6-10 mV and the tip was stopped every 15-25 pm. In each approaching-separating cycle, 50-100 I-V traces are acquired. **c,** Experimental I-V curves showing the voltage offset due to the temperature difference.

## 3. Thermal circuit

By heating the tip we do not only establish a temperature difference between the tip and the substrate but also a temperature gradient across the tip-connecting lead, which gives rise to an additional thermoelectric voltage. The equivalent circuit for measuring the thermopower is shown in Fig. S3. Considering the equivalent circuit, we can write:

$$\frac{I}{G} = V_{bias} - S(T_h - T_c) - S_{lead}(T_c - T_h) = V_{bias} - S\Delta T + S_{lead}\Delta T, \tag{S1}$$

where $V_{bias}$ is the bias voltage, $S$ is the thermopower of the junction, $S_{lead}$ is the thermopower of the tip-connecting lead, $T_c$ is the temperature of the substrate and $T_h = T_c + \Delta T$ is the temperature of the tip. The tip-connecting lead is a copper wire, so $S_{lead} = S_{Cu} = 1.83\ \mu V/K$ [S1]

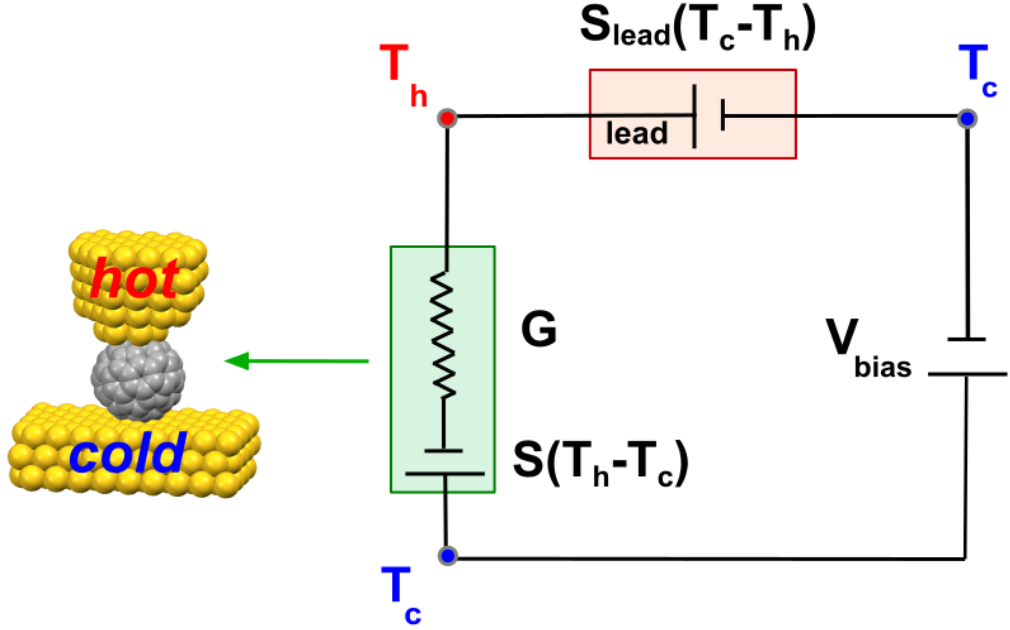

**Figure S3 | Equivalent thermal circuit of the setup for the determination of the thermopower.** The substrate and body of the STM are at ambient temperature $T_c$ while the tip is heated to a temperature $T_h = T_c + \Delta T$ above ambient temperature. $S$ is the thermopower of the molecular junction and $S_{lead}$ is the thermopower of the tip-connecting lead. $V_{bias}$ is the bias voltage applied to the junction.

From equation (S1) and for $I = 0$, the temperature-dependent voltage offset $V_{th}$ of the I-V curve is given by:

$$V_{th} = S\Delta T - S_{lead}\Delta T = (S - S_{lead})\Delta T. \tag{S2}$$

# 4. Pressure variation with tip displacement of the conductance, the thermopower, and the power factor of $Sc_3N@C_{80}$ junctions

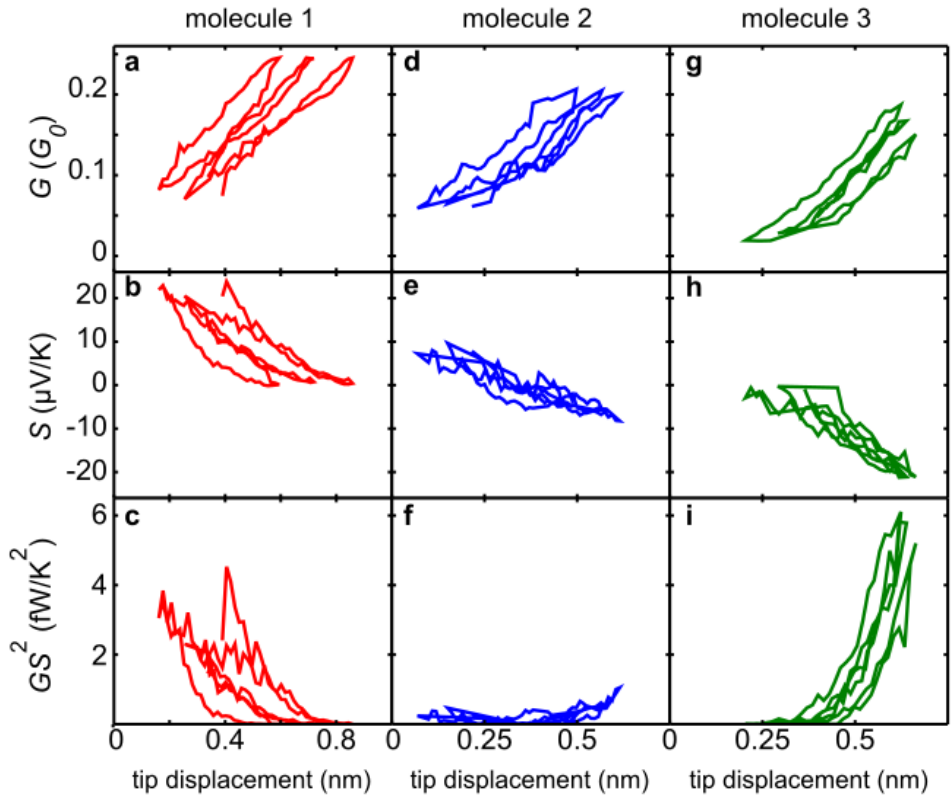

**Figure S4 | Effect of pressure on $Sc_3N@C_{80}$ junctions as a function of tip displacement.** Periodical variations of the conductance $G$ (**a,d,g**), the thermopower $S$ (**b,e,h**), and power factor $GS^2$ (**c,f,i**) as a function of the tip displacement. This is the same data as in Fig. 3a-c and the colours correspond to same three molecules detailed in the main text (1, 2 and 3). For molecule 1 (red), compressing the molecule results in $S$ varying from +20 µV/K to almost 0 µV/K; for molecule 2 (blue), $S$ varies from +10 µV/K to -5 µV/K; and for molecule 3 (green), $S$ varies from approximately 0 µV/K to -20 µV/K. Each half cycle corresponds to less than 0.5 nm. This representation is directly equivalent to the plot of the theoretical calculations in Fig. 4d-f.

## 5. Abrupt changes in the sign of the thermopower during junction formation

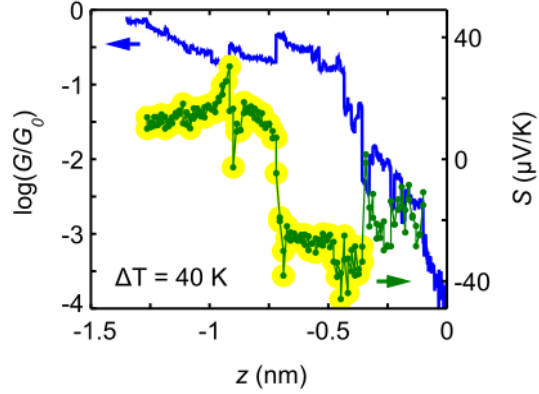

**Figure S5 | Example of an abrupt change in the sign of the thermopower during junction formation.** At the contact-point ($z = -0.35$ nm), the molecule shows negative Seebeck coefficient $S = -30$ µV/K and conductance $G = 0.2\ G_0$. After moving the tip closer to the surface by $\Delta z = 0.3$ nm (pressing the molecule), a sudden modification of the junction takes place and we observe a drop in conductance and an inversion of the sign of $S$: it flips from negative to positive ($S = +15$ µV/K).

## 6. Binding energy of $Sc_3N@C_{80}$ on a gold surface

To calculate the optimum binding distance for a $Sc_3N@C_{80}$ molecule between two gold (111) surfaces we use DFT and the counterpoise method, which removes basis set superposition errors (BSSE). The binding distance $z$ is defined as the distance between the gold surface and the $C_{80}$ cage at the closest point. Here the $Sc_3N@C_{80}$ molecule is defined as monomer A and the gold electrodes as monomer B. The binding distance $z$ is calculated for two different orientations for the molecule, the angle $\theta$ defines the orientation of the $Sc_3N$ molecule within the cage with respect to the gold surface. $\theta = 0°$ corresponds to the plane of this molecule being perpendicular to the gold surface (Fig. S6 top left) and $\theta = 90°$ (Fig. S6 top right) is when the plane of the $Sc_3N$ molecule lies parallel to the gold surface.

The ground state energy of the total system is calculated using SIESTA and is denoted $E_{AB}^{AB}$, with the parameters defined as those in the method section of the main text. Here the gold leads consist of 3 layers of 25 atoms. The energy of each monomer is then calculated in a fixed basis, which is achieved through the use of ghost atoms in SIESTA. Hence the energy of the individual $Sc_3N@C_{80}$ in the presence of the fixed basis is defined as $E_A^{AB}$ and for the gold is $E_B^{AB}$. The binding energy is then calculated using the following equation:

$$\text{Binding Energy} = E_{AB}^{AB} - E_A^{AB} - E_B^{AB} \tag{S3}$$

Fig. S6 shows that for an orientation $\theta = 0°$ the optimum binding distance $z$ is 2.5 Å and the binding energy is approximately 1.6 eV. For an angle of $\theta = 90°$ the value of $z$ is approximately 2.4 Å and has a binding energy of 1.8 eV.

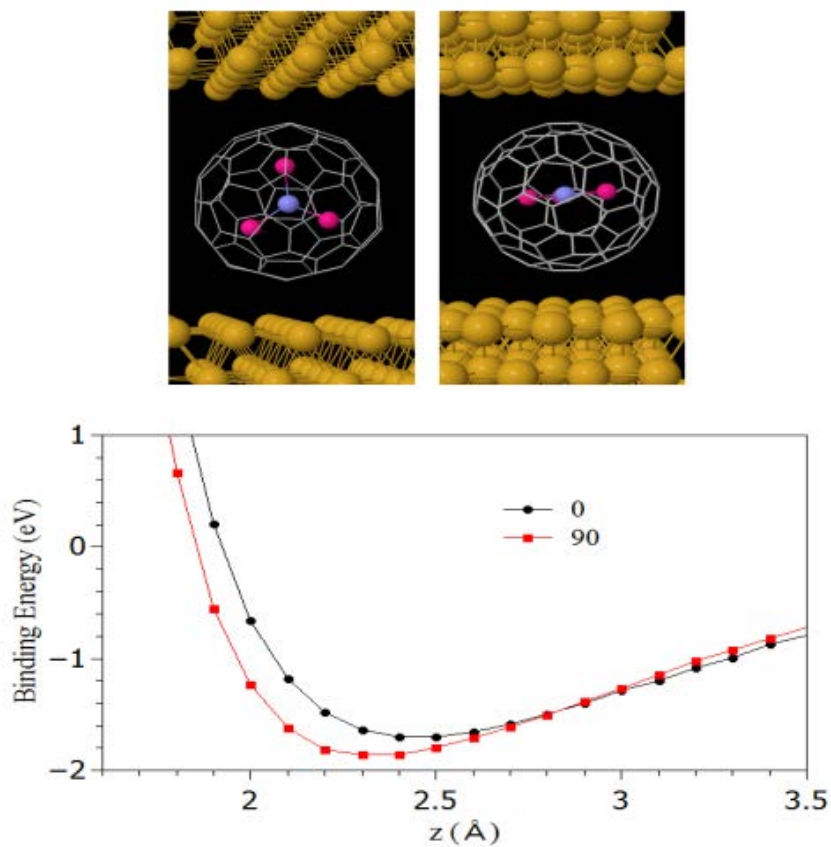

**Figure S6 | Sc$_3$N@C$_{80}$ on a gold surface**. Top pannels: Orientation of the Sc$_3$N@C$_{80}$ molecule with respect to the gold leads corresponds to the defined angle (left) $\theta = 0°$ and (right) $\theta = 90°$. Lower pannel: Binding energy of Sc$_3$N@C$_{80}$ to gold as a function of molecule-contact distance. The equilibrium distance is found to be approximately 2.3 Å from the minimization of the binding energy.

## 7. Frontier orbitals of the Sc₃N@C₈₀ molecule

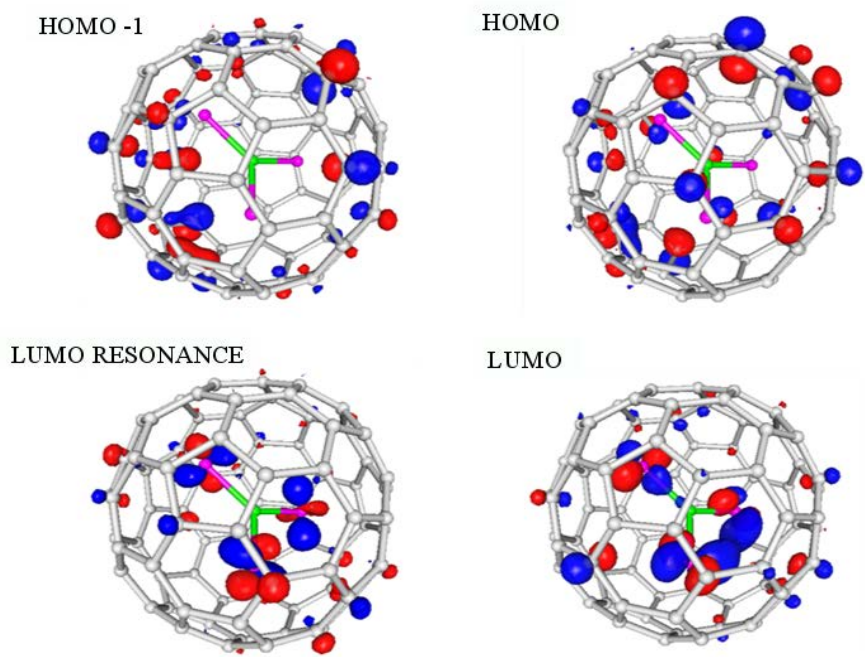

**Figure S7 | Local density of states of the HOMO-LUMO orbitals of the Sc$_3$N@C$_{80}$ molecule.**

## 8. Binding energy for Sc₃N within Ih-C₈₀ cage

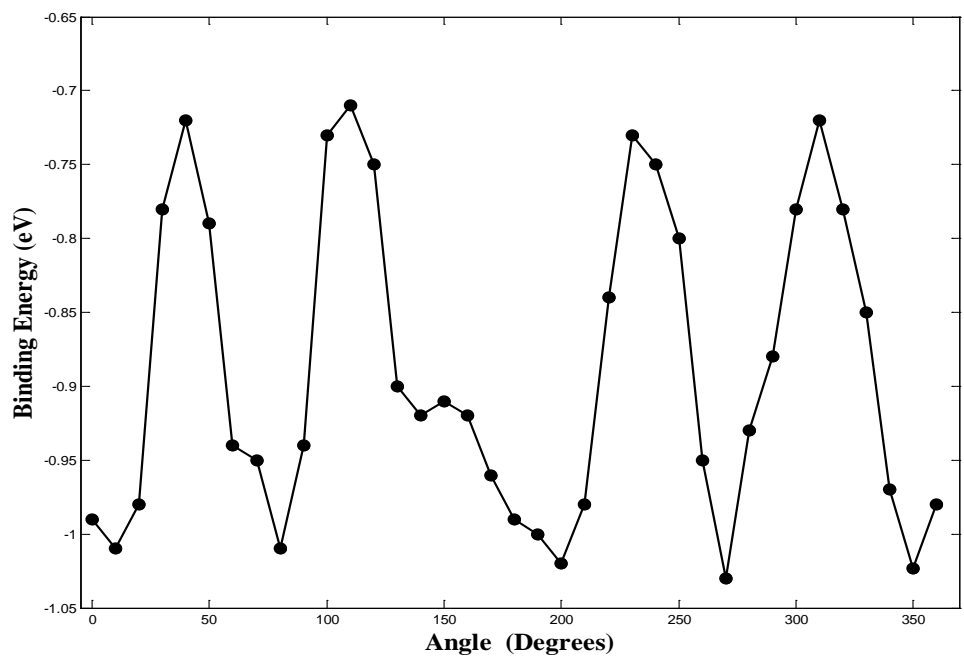

**Figure S8 | Binding energy as a function of rotation of Sc$_3$N within the fixed Ih-C$_{80}$ cage.** The energy barrier to rotation is of the order of 300 meV and therefore the Sc$_3$N is locked n position within the fullerene cage at room temperature.

## 9. Transmission coefficient as a function of orientation of the Sc$_3$N@C$_{80}$ molecule

Fig. S6 shows the definition of the orientation angle of the Sc$_3$N molecule for $\theta = 0°$ and $90°$. Fig. S9 shows the indivial transmission coefficients $\mathcal{T}(E)$ for various angles as the molecule is rotated through 180°, for indentical tip separations of $z = 2.3$ Å at each electrode. The DFT-predicted Fermi energy $E_F^0$ is close to the LUMO resonance and the effect of rotation causes this peak to shift and broaden. In the case of $\theta = 90°$ the resonance is narrowest and shifted furthest to the left. This is because at this orientation the scandium atoms are furthest from the gold surfaces and as the LUMO is located mainly on these atoms (Fig. S7) the coupling to the leads is weakest for this geometry.

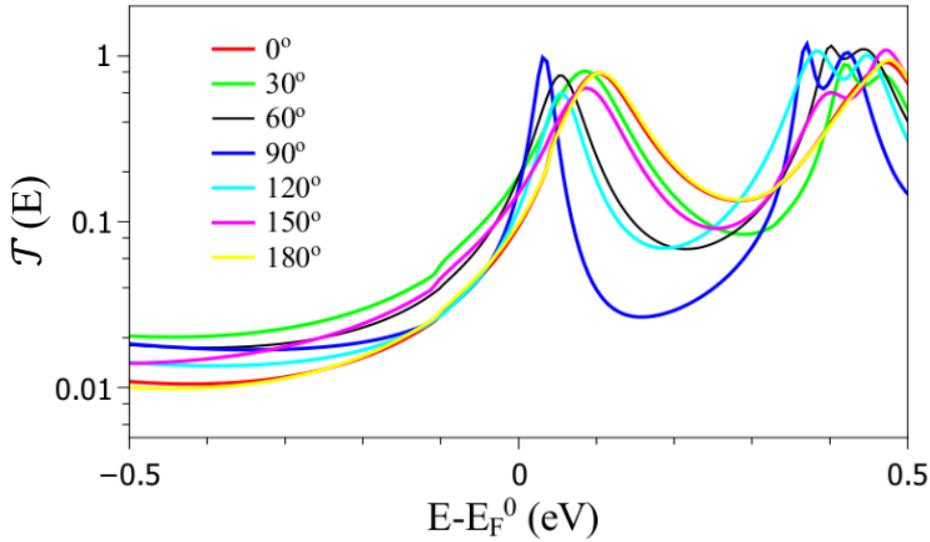

**Figure S9 | Transmission coefficient as a function of orientation**. Zero bias transmission coefficient $\mathcal{T}(E)$ versus electron energy $E$ for rotation angles $\theta$ between 0° and 180° of the Sc$_3$N@C$_{80}$ molecule with respect to the gold leads.

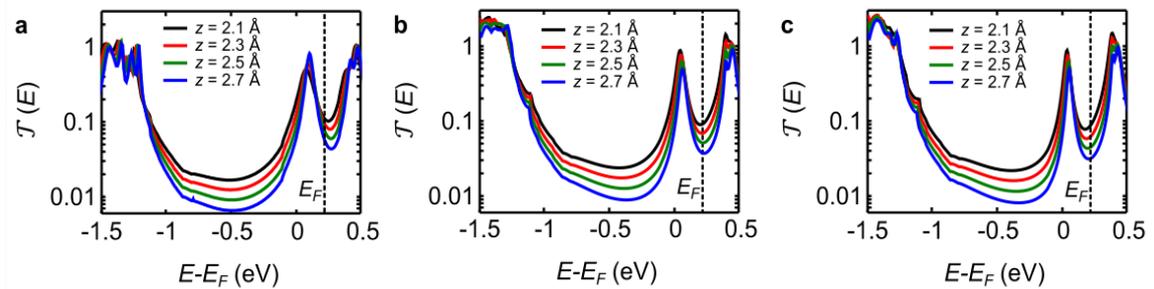

**Figure S10 | Transmission coefficient as a function of orientation and tip separation**. Transmission curves, $\mathcal{T}(E)$, for three different orientations ($\theta = 150°$, $\theta = 57°$, and $\theta = 63°$, respectively) and for different $z$. The data are the same as in Fig. 3g-i, but covering a wider energy range.

## 10. Calculated thermopower as a function of orientation and tip separation

To calculate the thermopower of these molecular junctions, it is useful to introduce the non-normalised probability distribution $P(E)$ defined by

$$P(E) = -\mathcal{T}(E)\frac{df(E)}{dE} \tag{S4}$$

where $f(E)$ is the Fermi-Dirac function and $\mathcal{T}(E)$ is the transmission coefficients and whose moments $L_i$ are denoted as follows

$$L_i = \int dE P(E)(E - E_F)^i \tag{S5}$$

where $E_F$ is the Fermi energy. The thermopower, $S$, is then given by

$$S(T) = -\frac{1}{eT}\frac{L_1}{L_0} \tag{S6}$$

where $e$ is the electronic charge.

Fig. S11 shows the thermopower $S$ evaluated at room temperature for orientation angles of $\theta$ from 0° to 180° and at four different tip separations $z$ from 2.7 to 2.1 Å. Here the value of the Fermi energy is $E_F = E_F^0 + 0.23$ eV, where $E_F^0$ is the DFT-predicted value of $E_F$. This value of $E_F$ has been optimized to give the best agreement with the experimental measurements. This plot shows how the behaviour of the thermopower is strongly dependent on the orientation of the $Sc_3N$ molecule within the cage. At low and high angles, i.e. less than 50° and greater than 130° the thermopower is positive, while between these angles it is negative. As the tip is moved closer to the molecule the value of the thermopower decreases at all angles, and at certain angles such as approximately 60° the sign of $S$ goes from positive to negative.

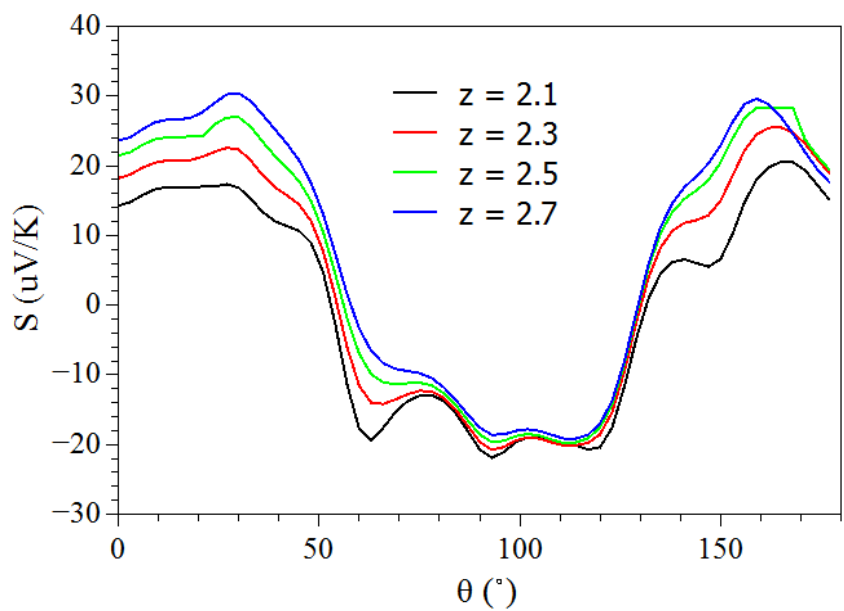

**Figure S11 | Thermopower as a function of orientation and tip separation.** Thermopower $S$ versus orientation angle $\theta$ at a value of $E_F = 0.23\ eV$, for different tip-substrate distances $z$.

## 11. Optimized geometries of Sc$_3$N@C$_{80}$ junctions

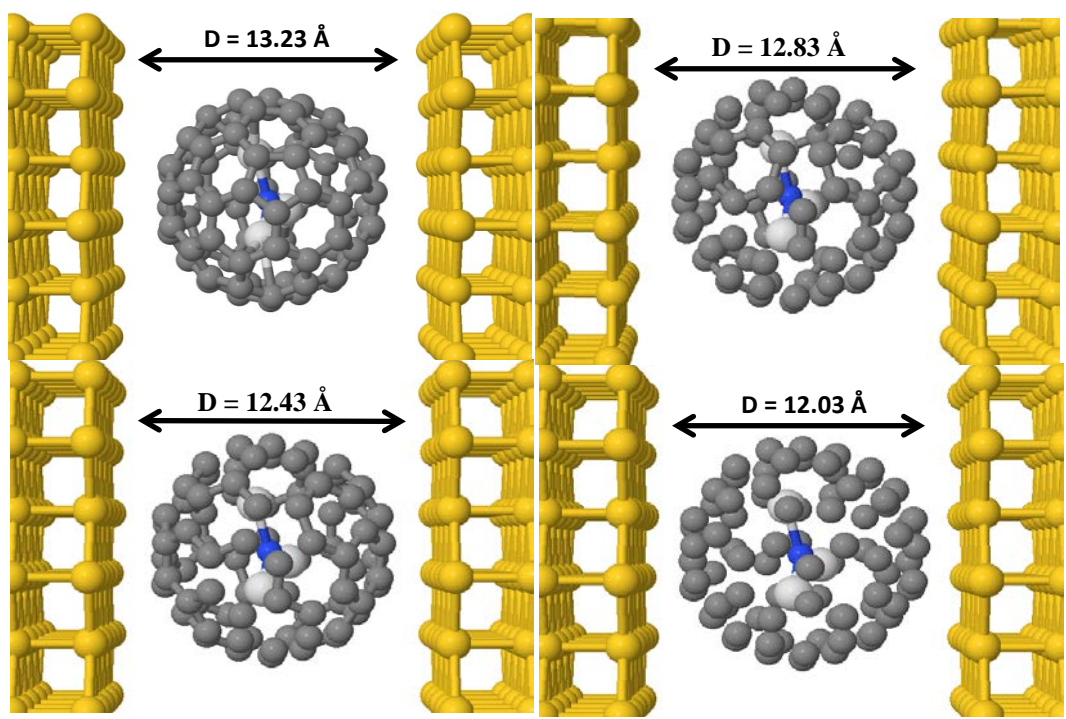

**Figure S12 | Optimized geometries of Sc$_3$N@C$_{80}$ junctions.** Four optimized geometries corresponding to tip-carbon distances of 2.5, 2.3, 2.1 and 1.9 Angstroms. They show that the C$_{80}$ barely distorts over such a range. In these simulations, the gold is not allowed to relax.